\documentclass[aps,a4paper,showkeys,manuscript]{revtex4}
\usepackage{amsmath}
\usepackage{color}%
\usepackage{graphicx}
\usepackage{epstopdf}
\usepackage{dcolumn}
\usepackage{bm}
\usepackage{bbding}
\usepackage{booktabs}            
\usepackage{amssymb,bm,mathrsfs,bbm,amscd} 
\usepackage{longtable}
\usepackage{mathrsfs}

\begin{document}
\title{Effects of cluster correlations on fragment emission in $^{12}$C + $^{12}$C at 50 MeV/nucleon}
\author{R. Han$^{1}$, Z. Chen$^{1,2}$\footnote{zqchen@impcas.ac.cn}, R. Wada$^{3}$\footnote{wada@comp.tamu.edu},
 A. Ono$^{4}$, G. Tian$^{1}$, F. Shi$^{1}$, X. Zhang$^{1,2}$, B. Liu$^{1,2}$, H. Sun$^{1,2}$}
\affiliation{$^{1}$Institute of Modern Physics, Chinese Academy of Sciences, Lanzhou, 730000, China}
\affiliation{$^{2}$University of Chinese Academy of Sciences, Beijing, 100049, China}
\affiliation{$^{3}$Cyclotron Institute, Texas A$\&$M University, College Station, TX 77843, United States}
\affiliation{$^{4}$Department of Physics, Tohoku University, Sendai 980-8578, Japan}
\begin{abstract}
The effects of cluster correlations have been studied in the $^{12}$C + $^{12}$C reaction at 50 MeV/nucleon, using three AMD models,
the AMD (AMD/D) without any additional cluster correlations,
AMD/D-COALS with nucleon correlations based on a coalescence prescription for light cluster formations with $A\leq4$
and AMD-Cluster with an extended cluster correlation in two-nucleon collision processes and a special treatment for intermediate fragment formation with A $\le 9$.
The angular distributions and energy spectra of fragments have been simulated and compared with the available experimental data.
It is found that the cluster correlations take a crucial role to describe the productions of light charged particles (LCPs) and intermediate mass fragments (IMFs),
and the AMD-cluster studied here provides a consistent overall reproduction of the experimental data.
It is also shown that the significant effects of the secondary decay processes are involved for the fragment production besides the dynamical productions in the AMD stage.
Detailed LCP and IMF production mechanisms involved in the intermediate energy heavy ion collisions are discussed.
\end{abstract}

\maketitle
\section*{I. Introduction}
The studies of the intermediate energy heavy-ion collisions are important to explore for nuclear matter properties and to understand nuclear reaction mechanisms.
In the central collisions, the system is compressed at an early stage and then expands so that the whole system disintegrates into many intermediate mass
fragments (IMFs) and light particles (LCPs). The process is very complicated under various conditions, such as densities, excitation energies, isospin asymmetries and so on.
During the past decades, a series of heavy-ion collisions have been performed and the double differential fragmentation cross sections have been measured~\cite{Divay2017,Dudouet2013,De2012,Bini2003,Hudan2003,Re2010,Ha2017}.
For example, the $^{12}$C + $^{12}$C reaction at 50 MeV/nucleon was performed at GANIL, motivated by the particle beam therapy
to treat cancerous tumors, and the experimental data are available in Ref.~\cite{Divay2017}.
In the carbon therapy treatment, accurate fragmentation cross sections in a wide energy range are necessary.
But the available experimental data are limited such as discontinuity in the incident energy range and limited target materials.
For the production rate of secondary particles, their angular and energy distributions are often not well known.

In order to elucidate the process of heavy-ion collision dynamics with the validated experimental data, many microscopic transport models have been developed.
Among these, the antisymmetrized molecular dynamics (AMD) model of Ono et al.~\cite{Ono1996,Ono1999}, in which the many-body nucleon wave function is antisymmetrized,
has achieved great success in describing many nuclear reaction phenomena for intermediate energy heavy-ion collisions~\cite{Ono2004,Wada2004,Ono2003,Ikeno20162,Piantelli2019}.
AMD not only solves the time evolution of many-nucleon system in a given mean field including quantum features,
but also can treat the cluster correlations in a stochastic manner.
Especially, as shown in our previous works in Refs.~\cite{Tian2017,Tian2018}, the cluster correlation has strong impacts on the whole collision dynamics,
not only for the formation of light clusters, but also for the production of heaver IMFs~\cite{Tian2017,Tian2018}.

The aims of this article are to examine different approaches for the cluster correlation, which refers to the processes involving clusters,
and to verify the importance of the cluster correlations in heavy-ion collisions.
We study the effects of the cluster correlations on fragment emissions in $^{12}$C + $^{12}$C at 50 MeV/nucleon using AMD models,
a version of AMD without any additional cluster correlations, and that with a cluster formation based on
a coalescence prescription for fragments with $A \leq 4$ and AMD-Cluster with an extended cluster correlation.
Particular interest will be focused on the production of LCPs and IMFs.

This paper is organized as follows. In Sec. II, the three versions of AMD used in the present work are briefly described.
In Sec. III, detailed comparisons between the experimental data and the different AMD model simulations are presented.
In Sec. IV, fragment production mechanisms and the sequential decay effects are discussed.
A summary is given in Sec. V.
\section*{II. Antisymmetrized Molecular Models}
\subsection*{A. AMD/D}
In the AMD model, the wave function for an $A$-nucleon system is described by a Slater determinant $|\Phi\rangle$,
\begin{equation}
|\Phi\rangle = \frac{1}{\sqrt{A!}}\det[\varphi_i(j)],
\end{equation}
where $\varphi_i=\phi_{Z_i}\chi_{a_i}$. The spin-isospin state $\chi_{a_i}$ of each single-particle state takes $p\uparrow, p\downarrow,n\uparrow$, and $n\downarrow$.
The spatial wave functions of nucleons $\phi_{Z_i}$ are given by a Gaussian wave function,
\begin{equation}
\langle{\bf r|\phi_{Z_i}}\rangle = \left(\frac{2\nu}{\pi}\right)^{3/4}\exp\left[-\nu\left({\bf r}-\frac{{\bf Z}_i}{\sqrt{\nu}}\right)^2+\frac{1}{2}{\bf Z}^2_i\right],
\end{equation}
where the width parameter $\nu=0.16$ fm$^{-2}$~\cite{Ono19921} is a constant parameter common to all the wave packets. Thus the complex variables
$Z\equiv\{{\bf Z}_i;i=1,...A\}=\{{Z}_{i\sigma};i=1,...A,\sigma=x,y,z\}$ represent the centroids of the wave packets.
Up to the antisymmetrization effect, the real part and the imaginary part of ${\bf Z}_i$ correspond to the centroids
of the position and the momentum, respectively,
\begin{equation}
{\bf Z}_i = \sqrt{\nu}{\bf D}_i+\frac{i}{2\hbar\sqrt{\nu}}{\bf K}_i,
\end{equation}
where {\bf D} = $\langle\phi_Z|{\bf r}|\phi_Z\rangle/\langle\phi_Z|\phi_Z\rangle, {\bf K} = \langle\phi_Z|{\bf p}|\phi_Z\rangle/\langle\phi_Z|\phi_Z\rangle.$
The AMD wave function $|\Phi\rangle$ contains many quantum features in it and well describes the ground states of nuclei.

The time evolution of the wave packet parameters $Z$ is determined by the time-dependent variational principle and the two-nucleon collision process.
The former is described as
\begin{equation}
\delta\int dt\frac{\langle\Phi(Z)|(i\hbar\frac{d}{dt}-H)|\Phi(Z)\rangle}{\langle\Phi(Z)|\Phi(Z)\rangle} = 0,
\end{equation}
The equation of motion for $Z$ derived from the time-dependent variational principle is
\begin{equation}
i\hbar\sum_{j\tau}C_{i\sigma,j\tau}\frac{dZ_{j\tau}}{dt} = \frac{\partial\mathcal{H}}{Z^{*}_{i\sigma}}.
\end{equation}
The matrix $C_{i\sigma,j\tau}$ ($i,j=1,2,\dots,A$ and $\sigma,\tau=x,y,z$) is a Hermitian matrix defined by
\begin{equation}
C_{i\sigma,j\tau} = \frac{\partial^2}{\partial Z^{*}_{i\sigma}\partial Z_{j\tau}}\log\langle\Phi(Z)|\Phi(Z)\rangle,
\end{equation}
and $\mathcal{H}$ is the expectation value of the Hamiltonian after the subtraction of the spurious kinetic energy of the zero-point
oscillation of the center of mass of fragments~\cite{Ono19921,Ono19922},
\begin{equation}
\mathcal{H}(Z) = \frac{\langle\Phi(Z)|H|\Phi(Z)\rangle}{\langle\Phi(Z)|\Phi(Z)\rangle}-\frac{3\hbar^2\nu}{2M}A+T_0[A-N_F(Z)],
\end{equation}
where $N_F(Z)$ is the fragment number, $T_0$ is $3\hbar^2\nu/2M$ in principle but treated as a free parameter for an overall adjustment of the binding energies.
The effective Hamiltonian in AMD is
\begin{equation}
H = \sum^A_{i=1}\frac{{\bf p}^2_i}{2M}+\sum_{i<j}\upsilon_{ij},
\end{equation}
where $M$ is the nucleon mass and $\upsilon_{ij}$ is the effective internucleon force. For the meanfield calculations, the effective interactions
such as Gogny force and Skyrme force have been usually employed in the Hamiltonian $H$. In this paper, the calculations with AMD/D is
performed with the standard Gogny force~\cite{Decharg1980}.

The wave packet parameter $Z$ do not have physical meaning when wave packets overlap with each other such as inside a nucleus
because of the effect of the antisymmetrization. Therefore physical coordinates W=$\{{\bf W}_i;i=1,\dots,A\}$ are defined approximately as
\begin{equation}
\mathbf{W}_i = \sqrt{\nu}{\bf R}_i+\frac{i}{2\hbar\sqrt{\nu}}{\bf P}_i = \sum^A_{j=1}(\sqrt{Q})_{kj}{\bf Z}_j,
\end{equation}
with
\begin{equation}
Q_{kj} = \frac{\partial \ln\langle\Phi(Z)|\Phi(Z)\rangle}{\partial({\bf Z}^*_k\cdot{\bf Z}^*_j)}.
\end{equation}

The NN collision process is treated as a stochastic process
using the above physical coordinates at each time step. The NN collision rate is determined by a given NN cross
section under Pauli principle. The NN cross section is given by~\cite{Ono2002}
\begin{equation}
\sigma(E,\rho) = \min \left(\sigma_{LM}(E,\rho),\frac{100\ \text{mb}}{1+E/(200\ \text{MeV})}\right),
\end{equation}
where $\sigma_{LM}(E,\rho)$ is the cross section given by Li and Machleidt~\cite{Li1993,Li1994}.
The angular distribution of proton-neutron scattering are parameterized as
\begin{equation}
\begin{split}
\frac{d\sigma_{pn}}{d\Omega}\propto 10^{-\alpha(\pi/2-|\theta-\pi/2|)}, \\
\alpha = \frac{2}{\pi}\max \{0.333\ln E[\text{MeV}]-1, 0\},
\end{split}
\end{equation}
while the proton-proton and neutron-neutron scatterings are assumed isotropic.

The dynamical effect of the quantum fluctuations in the Gaussian wave packet is treated
in the diffusion (and shrinking) process in the time evolution of the nucleon propagation~\cite{Ono1999,Ono2002}.
As described in details in the references, this process is taken into account in order to treat properly the multifragmentation process~\cite{Ono1999,Ono2002}.
In the present simulations, the version in Ref.~\cite{Ono1999} is used and it is called AMD/D in this article.

\subsection*{B. Coalescence treatment in AMD/D-COALS}
In order to improve the reproduction of the experimental data for the LCP yields,
a stochastic coalescence process~\cite{Ono2004,Ono2000} is introduced in AMD/D to take into account nucleon-nucleon correlations among N nucleons with $N\leq4$ under Pauli principle.
This treatment is referred as coalescence treatment in this article and the program is called AMD/D-COALS. The basic idea~\cite{Ono2000} is the following.
For each subsystem with a proton and a neutron, for example this stochastic process moves the centroids ${\bf Z}_a$ and ${\bf Z}_b$ to the same point $\frac{1}{2}({\bf Z}_a+{\bf Z}_b)$.
If the coalescence takes place, the deuteron probability will increase from $P_d=|\langle d|{\bf Z}_a-{\bf Z}_b\rangle|^2$ to 1.
The rate of the coalescence $c$ is determined by the requirement that the probability to find the two nucleons in the deuteron state $|d\rangle$ should be independent of time on average,
\begin{equation}
\frac{d}{dt}P_d+(1-P_d)c=0.
\end{equation}

This modification for the coalescence is necessary not only for a pair of a proton and a neutron
but also for pairs of light clusters when they form a single loosely bound intrinsic state.
In the present study, the following processes are taken into account,
$p+n\rightarrow d$, $p+n+n\rightarrow t$, $p+p+n\rightarrow ^{3}$He and $p+p+n+n\rightarrow \alpha$.
The coalescence rates are obtained in a similar way to the deuteron case so that the ground state probability of each subsystem is kept constant.
The coalescence should be considered only when no other nucleons will interact with the subsystem any more.
In order to take into account this condition, the coalescence is considered only for the subsystems in the low phase space density region of $f<1.5$,
where $f$ is a smeared phase-space occupation probability summed over the nucleon spin and isospin.

The coalescence process is a supplemental process for a transport model, on top of the mean-field propagation and a usual treatment of two-nucleon collisions.
With the stochastic coalescence treatment the nucleons in a formed cluster tend to move together in the mean field after the cluster has been formed.
However each nucleon are still treated as independent nucleons so that the cluster breaks up when a nucleon-nucleon collision occurs
for any of the constituent nucleons with one of nearby nucleons.
This calculation is called AMD/D-COALS in this paper, and the same NN cross section and Gogny interaction are used as those of AMD/D described in Sec. II.A.
\subsection*{C. Extended cluster correlation in AMD-Cluster}
The extended version of AMD is developed mainly to improve the description of the IMF emission by taking into account the cluster correlation
as a stochastic process of cluster binding.
This version with the extended cluster correlations has been called AMD-Cluster~\cite{Tian2018}.
As mentioned in section II.A, a two-nucleon collision process is introduced as a stochastic quantum branching process. This collision process can be treated as a transition
of an AMD wave function $|\Psi_i\rangle$ to one of the possible final states
$|\Psi_f\rangle$. The usual treatment of two-nucleon collisions is performed under the assumption that these two nucleons are not correlated with the other nucleons,
and that only the states of the scattered two nucleons are changed in the final states $|\Psi_f\rangle$.
The transition to a final state $|\Psi_f\rangle$ is assumed to occur instantaneously and to conserve the energy expectation value
$\langle\Psi_i|H|\Psi_i\rangle=\langle\Psi_f|H|\Psi_f\rangle$.
The residual interaction induces two-nucleon collisions and the transition rate is given by
\begin{equation}
W_{i\leftrightarrow f} = \frac{2\pi}{\hbar}|\langle\Psi_f|V|\Psi_i\rangle|^2\delta(E_f-E_i).
\end{equation}
For the AMD/D treatment, a density of states of the final configurations was applied by including the scattered state contribution of the two nucleons.
In AMD-Cluster, on the other hand,
the two-nucleon collision process allow the possibility that each colliding nucleon may form a cluster of mass numbers
$A$ up to 4 with some other wave packets. When two nucleons $N_1$ and $N_2$ collide, the process considered in general is
\begin{equation}
N_1+N_2+B_1+B_2\rightarrow C_1+C_2,
\end{equation}
in which each of the scattered nucleons $N_j (j = 1,2)$ may form a cluster $C_j$, which can be up to $\alpha$, with a spectator particle $B_j$.
The transition rate of the cluster-forming process is given by Fermi$^{\prime}$s golden rule. The transition rate is given by
\begin{equation}
\upsilon_{ij}d\sigma = \frac{2\pi}{\hbar}|\langle\varphi^{\prime}_1|\varphi^q_1\rangle|^2|\langle\varphi^{\prime}_2|\varphi^q_2\rangle|^2
|\mathcal{M}|^2\delta(E_f-E_i)\frac{p^2_{rel}dp_{rel}d\Omega}{(2\pi\hbar)^3}
\end{equation}
where $\mathcal{M}$ is the matrix element for the two-nucleon scattering to the final state with the relative momentum $p_{rel}$ and
the scattering angle $\Omega$ in the two-nucleon center-of-mass system.
where $|\varphi^{\pm q}_j\rangle = e^{\pm i{\bf q\cdot r}_j}|\varphi_j\rangle$ are the states after the momentum transfer $\pm {\bf q}$
to the nucleons $N_j (j=1,2)$, and $(p_{rel},\Omega)$ is the relative momentum between $N_1$ and $N_2$ in these states.
The above equation for the rate should be generalized because there are many possible ways of forming a clusters for each N of the scattered nucleons
$N_1$ and $N_2$. It should be done carefully. The method was extended in an interactive way to consider the formation of clusters
$C_1$ and $C_2$ with mass number $A$ up to 4. The general formula is
\begin{equation}
\label{18}
\begin{split}
\upsilon_{ij}d\sigma(C_1,C_2,p_{rel},\Omega) =& \frac{2\pi}{\hbar}P_1(C_1,p_{rel},\Omega)P_2(C_2,p_{rel},\Omega)|\mathcal{M}|^2\\
&\delta(E_f(C_1,C_2,p_{rel},\Omega)-E_i)\frac{p^2_{rel}dp_{rel}d\Omega}{(2\pi\hbar)^3}
\end{split}
\end{equation}
where the overlap probabilities in Eq. (\ref{18}) have been replaced by the probabilities of specific channels which satisfy
$\sum_{C_1}P_1(C_1,p_{rel},\Omega)=1$ and $\sum_{C_2}P_2(C_2,p_{rel},\Omega)=1$. The relative momentum $p_{rel}$ in the final state is
adjusted for the energy conservation depending on the channel $(C_1,C_2)$. The phase-space factor $p_{rel}^{2}/(\partial E_f/\partial p_{rel})$
also depends on $(C_1,C_2)$.

A more detailed description about AMD-Cluster is given in Ref. [13,16,25-27].
Here we only highlight how IMFs are formed in AMD-Cluster.
The basic procedure in the actual calculation is similar to the coalescence method applied for nucleons in AMD/D-COALS in the previous subsection,
but here the method is applied for heavier clusters with additional care and with some simplifications.
When moderately separated clusters ($1 < R_{rel} < 5$ fm) are moving away from each other with a small relative kinetic energy
({\bf R}$_{rel} \cdot$ {\bf V}$_{rel} > 0$ and $\frac{1}{2}\mu V^2_{rel}<12$ MeV where $\mu$ is the reduced mass),
their momenta are replaced by the center-of-mass momentum of two clusters.
The method has been introduced in Ref.~\cite{Tian2018}.
In addition to these conditions, linking is allowed only if each of the two clusters is one of the four closest clusters
of the other when the distance is measured by [({\bf R}$_{rel}$/3 fm)$^2$ + ({\bf V}$_{rel}$/0.25c)$^2$]$^{1/2}$.
It is further required one (or both) of the two clusters is an $\alpha$ cluster or is in a light nucleus already bound at
a previous time. Two clusters in different already-bound light nuclei are not linked. Nonclustered nucleons are treated here
in the same way as clusters but two nucleons are not allowed to be linked directly. Two clusters also should not be linked
directly if they can form an $\alpha$ or lighter cluster due to the combination of their spins and isospins. It is possible that more
than two clusters are eventually linked by these conditions. However, the process is canceled unless the mass of the linked
system is in the range 6 $\leq$ A $\leq$ 9, and therefore the binding usually occurs in dilute environment. The binding is performed
for the linked system by eliminating the momenta of clusters in the center-of-mass frame of the linked system.
The energy conservation should be achieved by scaling the relative radial momentum between the center-of-mass of the linked system
and a third particle. We choose a particle (a cluster or a non-clustered nucleon) that has the minimal value of
\begin{equation}
(r+7.5 \rm fm)(1.2-cos\theta)/min(\epsilon_{\|},5 MeV),
\end{equation}
where $r$ and $\epsilon_{\|}$ are the distance and the radial component of the kinetic energy for the relative motion between the linked system and
the third particle.
The factor with the angle $\theta$ between the relative coordinate ({\bf r}) and velocity ({\bf v}) is introduced so as to favor the case of
{\bf r}$\|${\bf v}. If the selected third particle is already in a bound light nucleus, the light nucleus is used as the third particle for the energy conservation.

AMD-Cluster simulations present in this article are performed, using the free NN cross section and the use of the different NN cross sections is discussed in Sec.VI.C.

\section*{III. Results}
In this section the comparisons between the simulated results and the experimental data are presented. The experimental data are taken from Ref.~\cite{Divay2017}.
The cluster correlation effects are studied in the simulations by three AMD models, AMD/D, AMD/D-COALS, and AMD-Cluster.
The calculations are performed in the impact parameter range of $b$=0-8 fm.
The AMD/D and AMD/D-COALS calculations are performed with the Gogny interaction~\cite{Decharg1980} and AMD-Cluster with the SLy4 interaction~\cite{EC1998}.
Terminologies "PLF" and "NN" components are also used same as those in Ref.~\cite{Tian2017}, where they stand for a projectile-like
and a nucleon-nucleon source component, respectively, defined by using a three moving source fit. The third component is a target-like fragment component.
Comparisons of the simulated results with the experimental ones are made both in the primary stage (AMD) and the secondary stages (AMD+GEMINI) as described below.
\section*{A. LCPs}
\begin{figure}[htbp]
\includegraphics [scale=0.75]{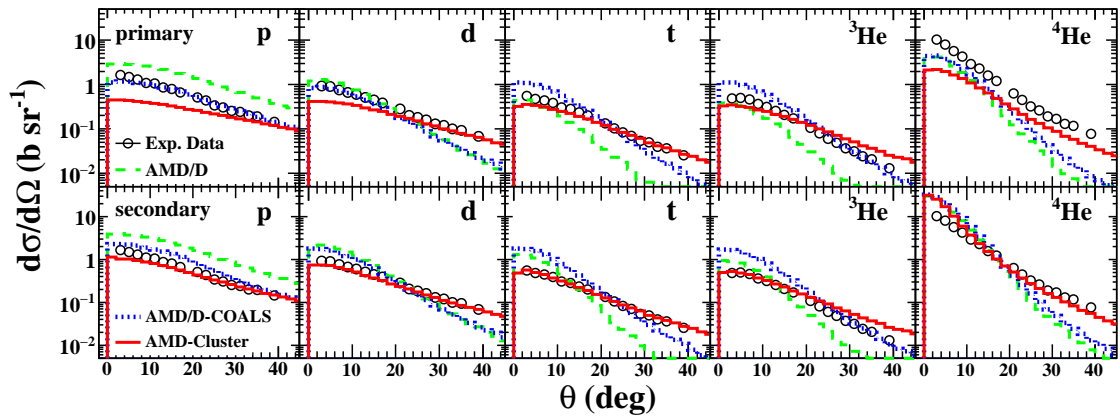}
\centering \caption{Comparisons of the angular distributions of the primary (upper row) and secondary (lower row) for p, d, t, $^3$He and $^4$He from left to right, respectively,
with different AMD simulations.
The experimental data, taken from Ref.~\cite{Divay2017}, are shown by open circles.
Green dash, blue dot and red solid lines show the simulated results from AMD/D, AMD/D-COALS and AMD-Cluster, respectively.} \label{fig1}
\end{figure}
On the upper row of Fig.~\ref{fig1}, the results of the primary angular distributions for different AMD simulations are shown
with the experimental data, though the latter are not directly compared to the primary yields.
A significant overprediction of protons at all angles are observed in AMD/D compared to AMD/D-COALS and AMD-Cluster.
The reduced yields of the latter two are very reasonable because the nucleon correlations, taken into account in the coalescence treatment and
the extended cluster correlation in AMD-Cluster, consume nucleons to form clusters.
For all LCPs, significant differences between the coalescence treatment and the AMD-Cluster treatment are observed.

On the lower row of Fig.~\ref{fig1}, the results of the secondary LCPs are shown in which GEMINI~\cite{Charity1988} is used as the afterburner.
(See APPENDIX for the difference between GEMINI++ which was used in our previous publications~\cite{Tian2017,Tian2018} and the Fortran Gemini which is used in this paper.)
AMD/D (green dash curves) significantly overtpredicts the yield of protons at all angles, inherited the trend in the primary yields.
For all other LCPs, it overpredicts at forward angles and significantly underpredicts at larger angles.
AMD/D-COALS (blue dot curves) improves the proton yield significantly, but still slightly overpredicts the yield at all angles.
For other LCPs, the trend is similar to those of AMD/D, but significantly overpredicts the PLF yields, especially for tritons and $^3$He.
In contrast to the former two calculations, AMD-Cluster (red solid curves) reproduces the angular distributions of all LCPs very well,
whereas a noticeable underprediction compared to the experimental data, is observed for protons at $\theta < 20^{\circ}$ and the significant overprediction are observed
for $^3$He at $\theta > 25^{\circ}$ and $^4$He at $\theta < 10^{\circ}$.
All three models significantly overpredict the experimental $^4$He yields by a factor of more than 2 at forward angles.
One should note that the significant overpredictions of the yields of $^4$He at the forward angles are dominated by the feeding from the secondary decays of PLFs,
which is revealed in the difference between those in the upper and lower rows of Fig.~\ref{fig1}.
No significant secondary contributions are observed at larger angles for all three calculations,
indicating that only a little additional feeding occurs at larger angles.

\begin{figure}[htbp]
\includegraphics [scale=0.7]{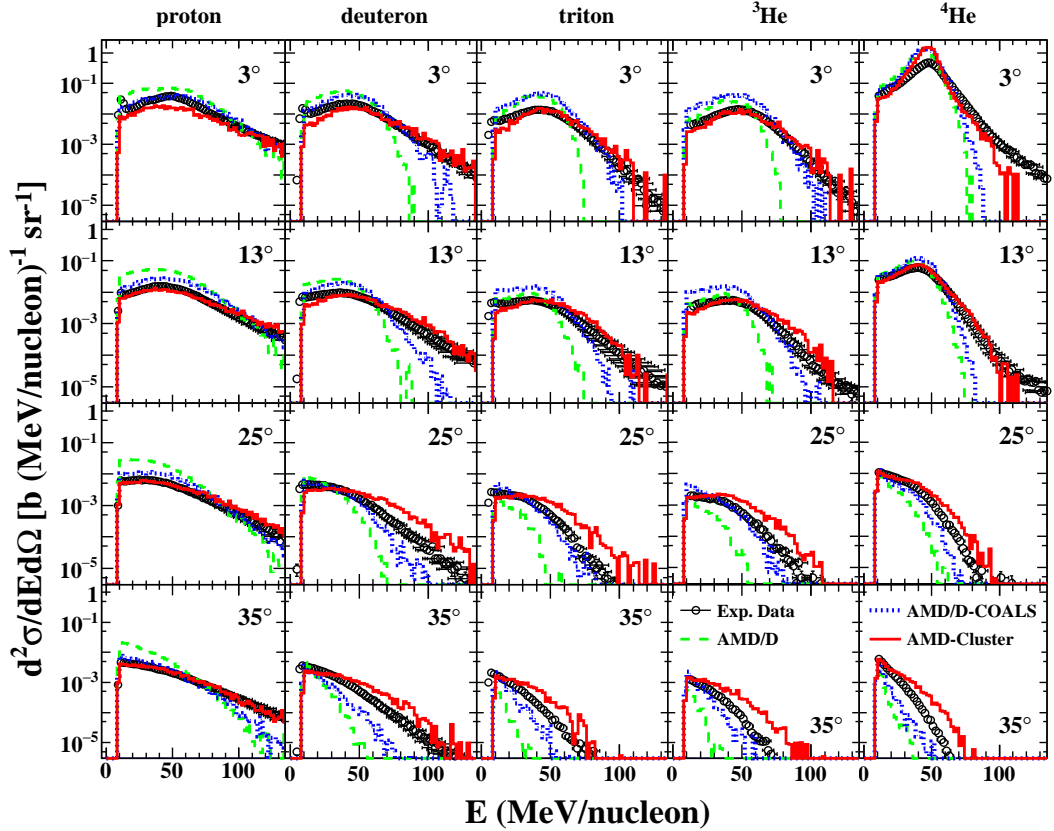}
\centering \caption{Comparisons of energy spectra of the secondary for p, d, t, $^3$He and $^4$He at selected angles.
See also the Fig. 1 caption for symbol and histograms.} \label{fig2}
\end{figure}
In Fig.~\ref{fig2}, energy spectra are compared between the experimental data and those of the secondary particles for LCPs.
\begin{enumerate}
\item[{\bf A1.}] {\bf Proton energy spectra}

The overprediction of protons in AMD/D, which is also shown in Fig.~\ref{fig1},
is observed in the entire energy range below the beam energy at all angles, and notable underestimation at the higher energies.
On the other hand AMD/D-COALS and AMD-Cluster improve the reproduction significantly in the entire energy range.
AMD/D-COALS starts to underpredict the proton yields at 35$^{\circ}$ and AMD-Cluster underpredicts the PLF yields at the most froward angle.

\item[{\bf A2.}] {\bf Deuteron energy spectra}

The overprediction of the PLF component in AMD/D is significantly improved both in AMD/D-COALS and AMD-Cluster, but the underpredictions at higher energies still remain in AMD/D-COALS, whereas notable overpredictions are observed in the AMD-Cluster at $\theta \geq 25^{\circ}$.

\item[{\bf A3.}] {\bf Triton and $^{3}$He energy spectra}

The overpredictions of AMD/D for the PLF component remains in AMD/D-COALS, but significantly improved in AMD-Cluster. On the other hand the significant underpredictions at higher energies are improved significantly in AMD/D-COALS, but AMD-Cluster overpredicts the yields at $\theta \geq 25^{\circ}$, where the feeding contributions are less, indicating that the primary yields of these clusters are overpredicted in AMD-Cluster.

\item[{\bf A4.}] {\bf Alphas energy spectra}

Alpha particles show overpredictions of the PLF yields in all three models.
The overprediction at forward angles generally originate from the feeding from the secondary decay of the PLFs.
\end{enumerate}
\section*{B. IMFs}
\subsection*{B.1 Angular distribution for Lithium and Beryllium isotopes}
\begin{figure}[htbp]
\includegraphics [scale=0.75]{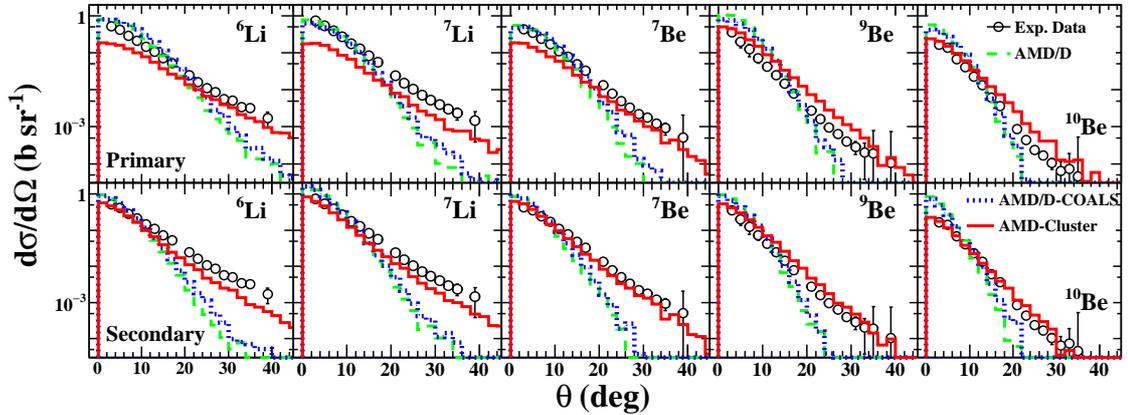}
\centering \caption{Comparisons of angular distributions of the primary (upper row) and secondary (lower row) for Lithium and Beryllium isotopes.
See also the Fig. 1 caption for symbol and histograms.
} \label{fig3}
\end{figure}

On the upper row of Fig.~\ref{fig3}, angular distributions of the primary Li and Be isotopes are shown.
As one can see, AMD/D and AMD/D-COALS show very similar results. This is reasonable because no cluster formation with $A>4$ is treated in an extended method in AMD/D-COALS.
Significant difference between these simulations and that of AMD-Cluster are observed.
AMD-Cluster tends to produce less IMFs at forward angles and more at larger angles.

On the lower row of Fig.~\ref{fig3}, the results of the secondary yields for these isotopes are shown.
For $^{6}$Li, $^{7}$Li and $^{7}$Be, AMD-Cluster reproduces the yields at forward angle in which feeding additions are dominated.
For $^{9}$Be and $^{10}$Be, the yields remain almost the same between the primary and secondary,
indicating the decay loss and the feeding addition from heavier IMFs are nearly balanced for these isotopes,
since as shown in Fig.8 below, most of the primary isotopes have the excitation energy much higher than the particle decay threshold
and only a few survive as the final products.
\subsection*{B.2 Angular distribution for Boron and Carbon isotopes}
\begin{figure}[htbp]
\includegraphics [scale=0.8]{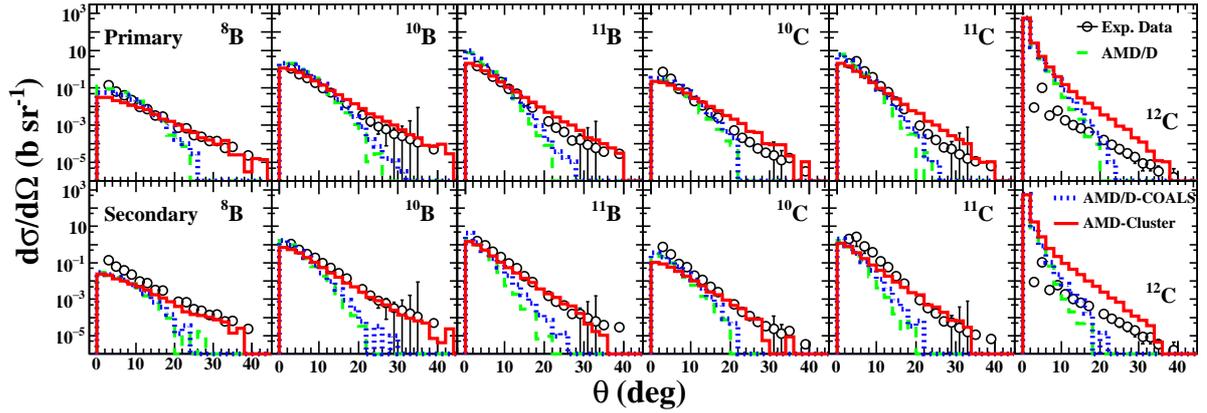}
\centering \caption{Comparisons of angular distributions of the primary (upper row) and secondary (lower row) for Boron and Carbon isotopes.
See also the Fig.~\ref{fig1} caption for symbol and histograms.
The large error bars in the original experimental data for $^{8}$B are not shown because they are dominated in the systematic errors of the detector calibration.
See details in Ref.~\cite{Tian2017}.} \label{fig4}
\end{figure}
In Fig.~\ref{fig4}, the angular distributions of the primary and secondary B and C isotopes are presented.
Compared to the primary, one can see the decay loss is slightly more than the feeding addition in the secondary yields for all three calculations.
For the secondary IMFs, a significant underpredictions are still observed for AMD/D and AMD/D-COALS at larger angles.
The yields of B and C isotopes at larger angles are rather well reproduced by AMD-Cluster,
though the yields at forward angles are slightly less compared to experimental data except for $^{12}$C.
For $^{12}$C, all three calculations show a similar trend as $^{4}$He results,
and significantly overpredict the yields at forward angles.
At the larger angles, the yields decrease sharply in the AMD/D and AMD/D-COALS and the experimental data is underpredicted, whereas AMD-Cluster still overpredicts significantly.

It is worth noting that distinct differences are observed in the angular distributions for C isotopes at 50 MeV/nucleon in this study and those at 95 MeV/nucleon in Refs.[15,16].
At 50 MeV/nucleon the experimental angular distribution for $^{12}$C shows more than one order less cross sections than those for $^{10}$C and $^{11}$C
and the angular distribution is much less forward peaking as shown in Fig.4. The angular distributions for $^{10}$C and $^{11}$C are similar each other in shape and amplitude.
On the other hand all AMD + Gemini simulations show comparable angular distributions for different C isotopes.
At 95 MeV/nucleon, both experimental and simulated angular distributions are very similar in shape and amplitude for different C isotopes
and the experimental data are well reproduced by AMD + GEMINI simulations.
It is not clear at this point what mechanism can produce such a drastic change in the experimental data between 50 MeV/nucleon and 95 MeV/nucleon for C isotopes.
Further studies are needed to elucidate this issue.

\subsection*{B.3 Energy spectra for IMFs}
\begin{figure}[htbp]
\includegraphics [scale=0.7]{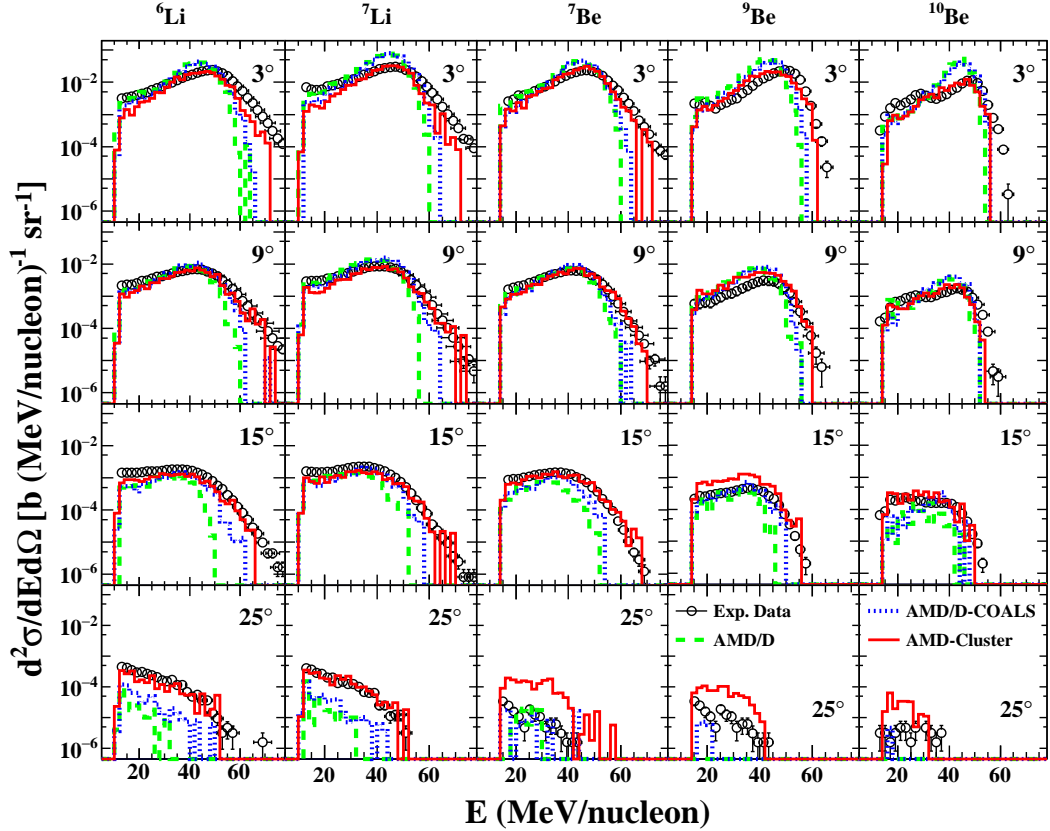}
\centering \caption{Comparisons of energy spectra of the secondary for Li and Be isotopes at selected angles.
See also the Fig.~\ref{fig1} caption for symbols and curves.} \label{fig5}
\end{figure}
In Fig.~\ref{fig5}, the comparisons of energy spectra are presented between the experimental data and those of the secondary yields for Li and Be isotopes.
In all cases AMD/D (green dashed lines) underpredict significantly the yields on the higher energy side
and AMD/D-COALS (blue dotted lines) are slightly better than AMD/D.
In AMD/D and AMD/D-COALS, the PLF components tend to be overpredicted by a factor of 1.5 to 3,
but both simulations reproduce the energy spectra below the beam velocity well for these light IMFs except for those at 25$^{\circ}$.
AMD-Cluster (red solid lines) reproduces the experimental data very well at entire energy range,
though the yields at 25$^{\circ}$ are significantly higher than to experimental data for Be isotopes.

\begin{figure}[htbp]
\includegraphics [scale=0.7]{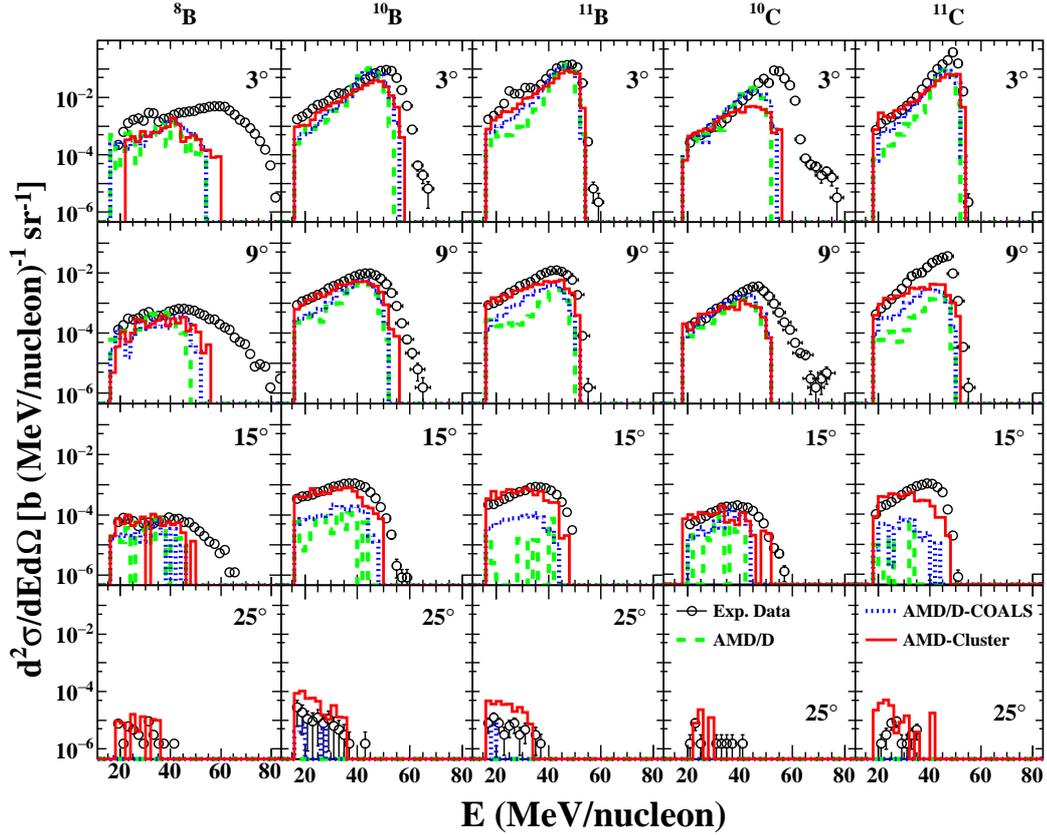}
\centering \caption{Similar plots as Fig.~\ref{fig5}, but for B and C isotopes.
See also the Fig.~\ref{fig1} caption.} \label{fig6}
\end{figure}
In Fig.6, the energy spectra of the secondaries for B and C isotopes are shown.
AMD-Cluster reproduces the experimental data reasonably well both in the angular distribution in Fig.4 and energy spectra in these figures
except for $^{8}$B and $^{10}$C at forward angles and on the higher energy side.
The observed discrepancies in $^{8}$B and $^{10}$C may originate in the experimental errors in the energy calibration for these isotopes,
since the PLF peak energies are too high, compared to those of the other isotopes. This may also be true for $^{10}$Be.
As mentioned earlier, the coalescence treatment in AMD/D-COALS is not applied for these clusters, whereas the treatment in AMD-Cluster includes these IMFs.
We will discuss this in more details in the next section.
\section*{IV. Discussions}
\subsection*{A. Production mechanisms of LCPs and IMFs}
Before we discuss the different production mechanisms in details, firstly we make a brief summary of the above observations.
\begin{enumerate}
\item Common results for all three models
\begin{enumerate}
\item[] $^{4}$He and $^{12}$C yields are significantly overpredicted at forward angles. (Figs.1, 2 and 4)
\end{enumerate}
\item Common results between AMD/D and AMD/D-COALS
\begin{enumerate}
\item[2.a] The PLF component of LCP clusters with $A \geq 3$ and light IMFs with $6 \leq A \leq 9$ in the energy spectra show significant overpredictions.
  (Figs.2 and 5)
\item[2.b] Angular distributions and energy spectra of IMFs with $A \geq 6$ are similar in these two simulations and these fragment yields are
significantly underpredicted at larger angles. (Figs.3, 4, 5 and 6)
\end{enumerate}
\item Common results between AMD/D-COALS and AMD-Cluster
\begin{enumerate}
\item[] Angular distributions and energy spectra of protons are well reproduced. (Figs.1 and 2)
\end{enumerate}
\item AMD-Cluster
\begin{enumerate}
\item[] The production of LCPs and IMFs are rather well reproduced in their angular distributions and energy spectra
except that $^{4}$He and $^{12}$C yields are significantly overpredicted at forward angles.
\end{enumerate}
\end{enumerate}

In the following, we discuss different production mechanisms for LCPs and IMFs by comparing the results in three models based on the above observations.
\begin{enumerate}
\item[a.] The point 1 relates to the overprediction of the primary yield and excitation energy of $^{12}$C at forward angles,
since the overprediction of $^{4}$He at forward angles originates in the feeding from the secondary decay of the excited PLFs.
Excited $^{12}$C in peripheral collisions is one of the main parent nuclei for the $^{4}$He production.
As shown in Fig.~\ref{fig4}, the yields of the PLF component of $^{12}$C are overpredicted significantly in all three calculations
and the angular distribution is similar to that of the secondary $^{4}$He particles.
\item[b.] The point 2.a is caused by the feeding from the excited PLF component of the IMFs,
suggesting that the excitation energy of these IMFs may be too high.
These LCP cluster yields are significantly affected by the feeding from the excited PLF component of the IMFs,
indicating that their parent nuclei are also excited too much.
We will discuss these issues in the next subsection.
\item[c.] The point 3 and 4 indicate the importance of the cluster correlations and IMF formation for the proton and all other cluster productions.
\end{enumerate}

The above observations reflect the different production treatments which are incorporated as different stochastic manners in the models.
The overprediction of protons in AMD/D is well treated by the coalescence process in the cluster correlations taken in other two models
and the IMF production is well reproduced by the cluster correlation and the IMF formation in AMD-Cluster.
On the other hand, the production of the PLF component of $^{12}$C is significantly overpredicted with too much excitation energies in all three models.
This may be closely related to the observation of the reverse kinematic collisions between $^{40,48}$Ca, $^{58,64}$Ni beams on $^{9}$Be target
at 140 MeV/nucleon in Ref. [31], which will be discussed in the next subsection.
Another production mechanism is the sequential secondary decay process.
This process is closely related to the primary AMD stage through the excitation energy given to IMFs.
This process not only alters the yields of IMFs by the decay loss, but also alters the yields of LCPs and IMFs by the feeding addition.
Therefore the primary AMD results are altered significantly by the secondary sequential decay processes as discussed further below.
The AMD-Cluster + GEMINI examined here provides overall consistent results in reproducing the experimental angular distributions
and energy spectra of LCPs and IMFs.
\subsection*{B. Excitation energy and Sequential decay}
\begin{figure}[htbp]
\includegraphics [scale=0.7]{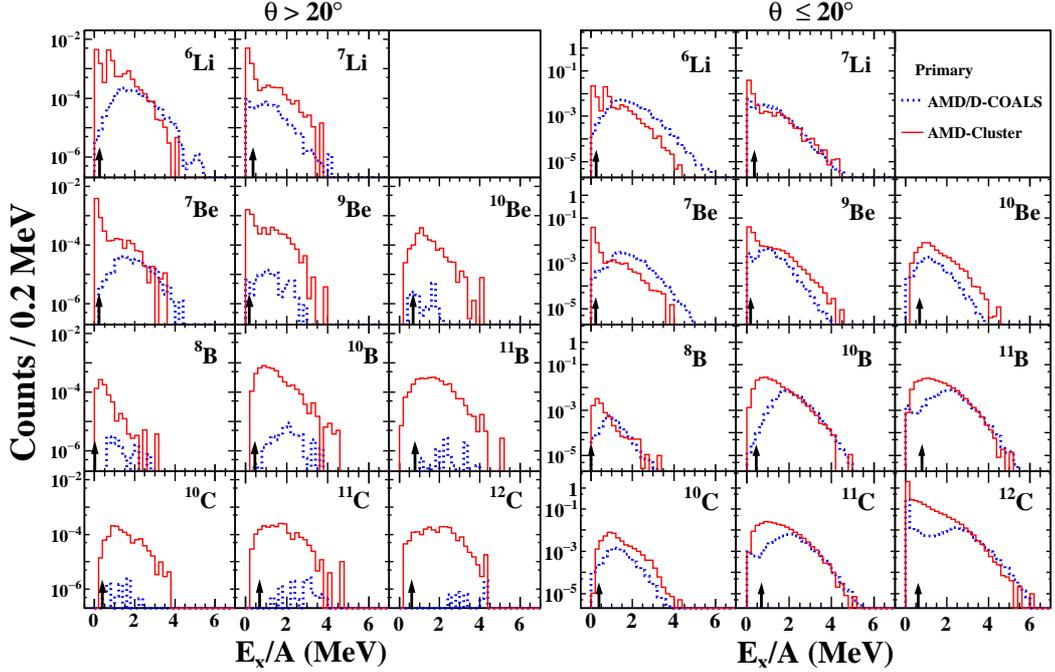}
\centering \caption{Excitation energy distributions of IMFs emitted at $\theta_{lab}>20^{\circ}$ (left panel) and $\theta_{lab}\leq20^{\circ}$ (right panel)
from AMD/D-COALS and AMD-Cluster and with kinetic energies above experimental energy thresholds.
Blue dot and red solid lines represent the excited primary fragment by the AMD/D-COALS and AMD-Cluster simulated, respectively.
The experimental particle decay thresholds are also shown on the X axis by black arrows.} \label{fig7}
\end{figure}
In this section we further study the excitation energy of IMFs given by AMD and sequential decay products generated by GEMINI in details.
On the left panel of Fig.~\ref{fig7}, the excitation energy distributions of IMFs observed at $\theta_{lab}>20^{\circ}$ from AMD/D-COALS and AMD-Cluster,
are compared for those with kinetic energies above the detection energy threshold in the experiment.
The same detection energy thresholds are applied to the primary fragments as those of the experiment.
This condition eliminates most of the TLF contribution to the distributions and therefore these IMFs belong mostly to the NN source component.
AMD/D shows very similar results to those of AMD/D-COALS, and therefore the results of AMD/D are not discussed in this section.
The excitation energy distributions of IMFs are quite different in AMD/D-COALS and AMD-Cluster
in two ways at this larger angle setup. One is the production yields and the other is the shape of the distributions.
The yields are significantly enhanced in the AMD-Cluster simulations,
indicating that AMD-Cluster efficiently produces the NN source component of the IMFs at these angles.
The second is that the cluster correlation generates the enhancement at the lower excitation energies.
These two facts result in the larger yields of the survived IMFs and of the feeding additions after GEMINI
at the larger angles observed in Figs. 3 to 6, which result in better reproductions of the experimental IMFs observations.

On the right panel of Fig.~\ref{fig7}, similar results are shown, but for $\theta_{lab}\leq20^{\circ}$,
and therefore the distributions are dominated by PLFs.
In this case the production yields are similar between those in AMD/D-COALS and AMD-Cluster,
but AMD-Cluster again produces a larger number of IMFs at lower excitation energies for the most isotopes except for $^{10}$Be,
and therefore a larger number of the survived IMFs after GEMINI.
However the difference in these survived IMFs is not observed in the secondary PLF distribution in Figs. 3 and 4,
since there are overwhelming contributions in the feeding process from the heavier IMF decays and they dominate the yields.
For the $^{12}$C case, there is a sharp peak near 0 MeV excitation energy in both simulations,
though AMD-Cluster shows much larger amount of the yields at low excitation energies.
These yields near 0 MeV (but those above the decay threshold) dominates the large enhancement of the alpha yields
from PLF fragments observed in the lower panels of Fig. 1 for $^{4}$He and Fig. 4 for $^{12}$C.
The excitation energy distributions of IMFs in AMD-Cluster are similar with our previous study for $^{12}$C+$^{12}$C at 95 MeV/nucleon~\cite{Tian2018},
indicating similar production mechanisms are involved in collisions between 50 MeV/nucleon and 95 MeV/nucleon.
There it is also pointed out that GEMINI predicts a notable amount of IMFs with the excitation energy above their particle threshold can survive as the final products.

According to Ref.~\cite{Mocko2008}, AMD/D tends to overpredict the excitation energy of PLFs, compared to that of the HIPSE
(heavy-ion phase-space exploration)~\cite{Lacroix2004} simulation
and causes significant overprediction of the IMF yields in the reverse kinematics reactions of $^{48}$Ca and $^{58,64}$Ni on $^{9}$Be at 140 MeV/nucleon
except for the $^{40}$Ca+$^{9}$Be system. In their analysis, the measured fragments show two distinct trends,
one is monotonically decrease from the projectile mass to lighter IMFs up to around A$\sim$15
and a rapid increase for the lighter IMFs. The latter yields are dominated by the multifragmentation events.
The former yields are contributed both from the primary production in the AMD stage and the feeding addition from the heavier IMFs in afterburner.
Therefore when we combine our results to their former IMF yields, their exceeding yields may be dominated from the exceeding production of
the projectile-like fragments and their secondary decay with too much excitation energy given to them.
However this scenario cannot be applied for $^{40}$Ca+$^{9}$Be system where the larger IMF yields are reasonably well reproduced by the AMD simulations.

In a recent application of AMD-Cluster for mid-peripheral and peripheral collisions of $^{93}$Nb+$^{93}$Nb and $^{116}$Sn at 38 MeV/nucleon in Ref.~\cite{Piantelli2019},
it shows that AMD-Cluster improved a lot in the fragments production for these peripheral collisions and reproduces the experimental data very well,
indicating that the excitation energy of PLFs is properly reproduced incorporating with the cluster correlation, which is consistent with our present results
except for the PLF component of the $^{12}$C. They also observed that Fortran version of GEMINI and C++ version result in noticeable differences
in the feeding production from the secondary decay process. However in our light reaction system,
we did not observed noticeable difference between these two versions after the proper treatment of the paring effect in the calculation, which is described in the Appendix A.

\subsection*{C. Effects of in-medium NN cross sections}
\begin{figure}[htbp]
\includegraphics [scale=0.7]{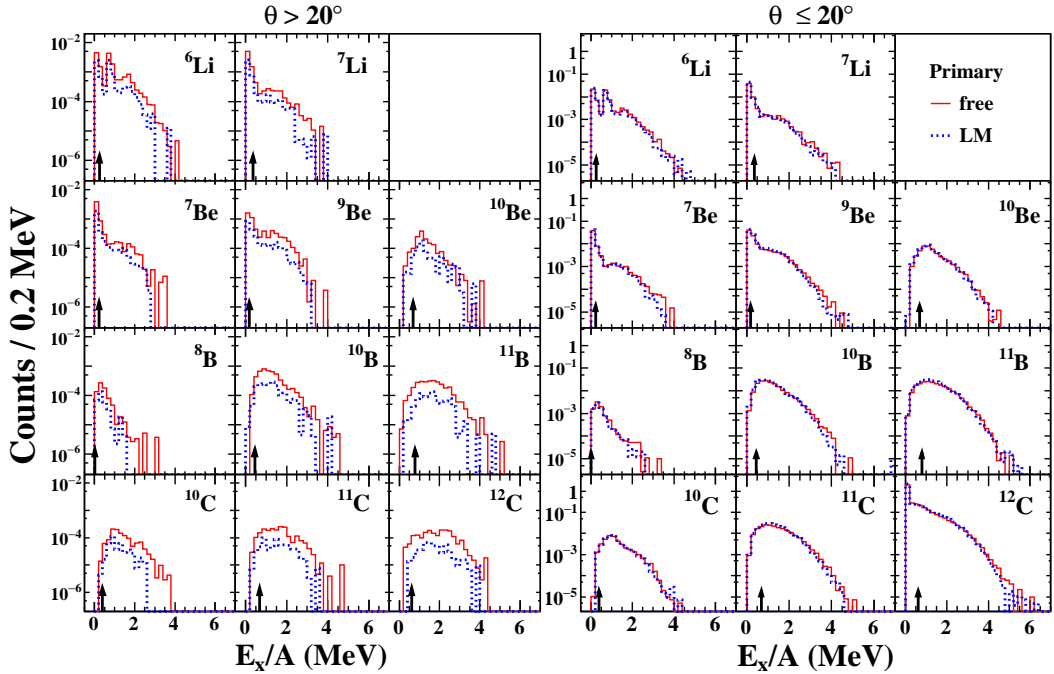}
\centering \caption{Similar plots as Fig.7, but for AMD-Cluster with two different NN cross sections.
Blue dot and red solid histograms represent LM and free NN cross sections, respectively.
See also the Fig.7 caption.} \label{fig8}
\end{figure}
AMD-Cluster reproduces the IMF angular distribution and energy spectra very well.
As discussed in the previous subsection, this improvement is made in two ways.
One is the enhanced IMF production in AMD and the second is feeding additions from the heavier IMF decays.
The latter is closely related to the excitation energy of the IMFs given in the AMD stage.
Since in AMD-Cluster, the cluster correlation is incorporated in two-nucleon collision process and it may be significantly affected by the in-medium NN cross section.
In order to verify this effect, the excitation energy distributions in two different NN cross sections are examined,
one is the free NN cross section shown in all above results in the AMD-Cluster simulations and the other is the LM cross section used in the above AMD/D and AMD/D-COALS simulations in Eq. (11).
The LM cross section is reduced roughly to $10\%$ at 10 MeV,
almost linearly increases to $35\%$ up to 50 MeV and gradually saturates up to $40\%$ for the higher energy compared to the free NN cross section~\cite{Lopez2014}.
The AMD-Cluster simulations are made with these two NN cross sections, but all other parameters are kept same.
The simulated excitation energy distributions are shown in Fig.8.
One can see that the distributions are almost identical at $\theta_{lab}\leq20^{\circ}$ on the right panel.
For those at $\theta_{lab}>20^{\circ}$ on the left panel, a noticeable reduction of the yields is observed for 6$\leq$A$\leq$12 isotopes in case of the LM cross section
compared to those with the free NN cross section. This reduction rate is similar to that of the NN cross section.
This observation indicates that the in-medium NN cross section does not take a dominant role
for the PLF production, but affects to the IMF production of the NN source component.
The significant improvement of the IMF production from the NN source shown at larger angles in Figs. 3 to 6 may be largely contributed by the use of the free NN cross section.

\section*{V. summary}
The effects of cluster correlations on the LCP and IMF productions are studied in the reaction $^{12}$C+$^{12}$C at 50 MeV/nucleon using three
AMD models, AMD/D, AMD/D-COALS and AMD-Cluster in which the same effective interaction is used in the former two models but with/without the cluster correlation
and in the latter two models cluster correlations are incorporated in different manners.
The angular distributions and energy spectra of all experimentally observed ejectiles have been compared between the
simulated results and the experimental data~\cite{Divay2017}.
The simulated AMD results are significantly altered by the sequential secondary decay of the excited IMFs.
For LCPs, the coalescence treatment in AMD/D-COALS improves the proton and deuteron yields very much compared to AMD/D,
but significantly overpredicts in the PLF component of triton, $^{3}$He and $^{4}$He yields.
AMD-Cluster reproduces the experimental data of LCPs and IMFs very well except for the PLF component of $^{4}$He and $^{12}$C.
For IMFs, the results from AMD/D and AMD/D-COALS are very similar and they significantly underpredict the yields for their NN component at larger angles.
For $^{4}$He and $^{12}$C, AMD/D and AMD/D-COALS significantly underpredict the yields at larger angles,
and furthermore, all three calculations significantly overpredict the yields at forward angles.
The AMD-Cluster simulations improve the PLF component of LCPs and the yields of IMFs at larger angles
compared to that of the AMD/D and AMD/D-COALS simulations.
The improvement of the IMF cluster production are made in two ways, one is the increased production rate of the primary IMFs
and the other is the proper amount of the excitation energy given to the primary IMF, including the enhanced population at the lower excitation energies.
The latter enhances the survival probability of the excited IMFs after the secondary decays,
though it is not directly observed because of the large contributions of the feeding additions.
In overall AMD-Cluster works best for this light reaction system in the intermediate energy range.
The cluster correlations take a crucial role to describe the production of LCPs and IMFs.

\section*{Acknowledgments}
This work is supported by the National Nature Science Foundation of China
(Grants Nos. U1832205, 11605257 and 11875298), and CAS "Light of West China" Program (Grant No. 29Y725030).
This work is also supported by the US Department of Energy under Grant No. DE-FG02-93ER40773 and the Robert
A. Welch Foundation under Grant A330. A. O. acknowledges support from Japan Society for the Promotion of Science KAKENHI Grant No. JP17K05432.
\appendix
\renewcommand{\appendixname}{Appendix}
\section{Fortran GEMINI and GEMINI++}
In our previous publications~\cite{Tian2018}, we used GEMINI++~\cite{Charity2010} and there is a slight difference with the Fortran GEMINI used in this work.
In GEMINI++, the available thermal energy U is evaluated slightly different from that in the Fortran GEMINI as
\begin{equation}
U = E^{*} - E_{yrast}(J) - \delta P,
\end{equation}
where $\delta$ P is the pairing correction to the empirical mass formula besides the shell correction.
No pairing correction is made in the Fortran Gemini. There is a control parameter Zshell for the correction.
The default value of Zshell=2, which means that the pairing correction is not apply for Z$\leq$2 clusters.
The paring correction for some of IMFs is set 5-10 MeV and increases the particle decay energy threshold and overpredict the IMF yields significantly as shown in Fig.3 of Ref.~\cite{Tian2018}.
When the particle decay is forced for IMFs with the excitation energy above the empirical particle energy threshold,
the yields are reduced by a reasonable amount and they become comparable to those of the experimental data.
Therefore we set Zshell=20 to turn off the pairing correction for IMF with Z$\leq$20, then results from GEMINI++ and Fortran-GEMINI become similar.
There are still some difference in the IMF excitation energy distributions, but no noticeable differences are observed in the multiplicity distributions.
Therefore we concluded that GEMINI++ with Zshell=20 and Fortran GEMINI end up in the essentially same results and in this work we used the Fortran GEMINI.

\end{document}